\documentclass[twocolumn,aps,prb,superscriptaddress]{revtex4-1}
\usepackage{graphicx}
\usepackage{float}
\usepackage{epstopdf}

\begin{document}

\title{Penetration depth study of LaOs$_4$Sb$_{12}$: Multiband $s$-wave superconductivity}
\author{X. Y. Tee}
\affiliation{Division of Physics and Applied Physics, School of Physical and Mathematical Sciences, Nanyang Technological University, Singapore 637371, Singapore}

\author{H. G. Luo}
\affiliation{Center for Interdisciplinary Studies and Key Laboratory for Magnetism and Magnetic Materials of the MoE, Lanzhou University,
Lanzhou 730000, China}

\author{T. Xiang}
\affiliation{Institute of Physics, Chinese Academy of Sciences, P. O. Box 603, Beijing 100190, China}
\affiliation{Institute of Theoretical Physics, Chinese Academy of Sciences, P. O. Box 2735, Beijing 100190, China}

\author{D. Vandervelde}
\affiliation{Department of Physics, University of Illinois at Urbana-Champaign, Urbana, Illinois 61801, USA}

\author{M. B. Salamon}
\affiliation{Department of Physics, University of Texas at Dallas, Richardson, Texas 75080, USA}

\author{H. Sugawara}
\affiliation{Faculty of Integrated Arts and Sciences, University of Tokushima, Tokushima 770-8502, Japan}

\author{H. Sato}
\affiliation{Department of Physics, Tokyo Metropolitan University, Tokyo 192-0397, Japan}

\author{C. Panagopoulos}
\affiliation{Division of Physics and Applied Physics, School of Physical and Mathematical Sciences, Nanyang Technological University, Singapore 637371, Singapore}

\author{Elbert E. M. Chia}
\affiliation{Division of Physics and Applied Physics, School of Physical and Mathematical Sciences, Nanyang Technological University, Singapore 637371, Singapore}

\date{\today}

\begin{abstract}
We measured the magnetic penetration depth $\lambda(T)$ in single crystals of LaOs$_{4}$Sb$_{12}$ ($T_c$=0.74 K) down to 85 mK using a tunnel diode oscillator technique. The observed low-temperature exponential dependence indicates a $s$-wave gap. Fitting the low temperature data to BCS $s$-wave expression gives the zero temperature gap value $\Delta (0)= (1.34 \pm 0.07) k_B T_c$ which is significantly smaller than the BCS value of 1.76$k_B T_c$. In addition, the normalized superfluid density $\rho(T)$ shows an unusually long suppresion near $T_c$, and are best fit by a two-band $s$-wave model.
\end{abstract}

\maketitle

The skutterudite PrOs$_4$Sb$_{12}$ has been receiving much attention due to the discovery of its unconventional heavy fermion superconductivity, and the possible role played by quadrupolar fluctuations in its pairing mechanism.~\cite{Bauer02,Maple02,Goto04,kuwahara05} An important step in clarifying its superconductivity is to determine the symmetry of the superconducting gap. Conflicting results had been reported. On one hand, scanning tunneling microscopy (STM), muon spin rotation ($\mu$SR), and nuclear quadrupolar relaxation (NQR) experiments point to an isotropic superconducting energy gap.~\cite{Suderow04,MacLaughlin02,Kotegawa03} On the other hand, magnetic penetration depth, angle-dependent thermal conductivity, and neutron scattering measurements revealed point nodes on the superconducting gap.~\cite{Elbert03,Izawa03,Huxley04} Angle-dependent thermal conductivity data revealed two distinct superconducting phases of twofold and fourfold symmetries, both with point nodes on the superconducting gap.~\cite{Izawa03} Recently, multiband superconductivity was proposed as a reconciliation for these conflicting results. According to field-dependent thermal conductivity data, PrOs$_4$Sb$_{12}$ is a two-band superconductor, with nodes on one band and an isotropic gap on the other.~\cite{Seyfarth05,Seyfarth06,Hill08} A similar two-gap structure was observed in Li$_{2}$Pt$_{3}$B, the nodes there arising from the interference of singlet and triplet pairing.~\cite{Yuan06}

LaOs$_{4}$Sb$_{12}$ (superconducting transition temperature $T_{c}$=0.74~K) and PrOs$_{4}$Sb$_{12}$ ($T_{c}$=1.85~K) are iso-structural superconductors. The substitution of La by Pr introduces the 4$f$ electrons, while largely preserving the crystal structure~\cite{Harima06,Sugawara02,Bauer01} and Fermi surface topology.~\cite{Sugawara02} The 4$f$ energy levels of the Pr$^{3+}$ ion are split by the crystal electric field (CEF), resulting in a non-magnetic $\Gamma_{1}$ singlet ground state, a 0.7-meV $\Gamma_{5}$ first excited state, and 11-meV $\Gamma_{4}$ second excited state (in $O_{h}$ cubic symmetry).~\cite{Goremychkin04} The higher $T_{c}$ of PrOs$_{4}$Sb$_{12}$ compared to LaOs$_{4}$Sb$_{12}$ is attributed to the \textit{type} of scattering between the conduction electrons and these low-lying CEF-split 4$f$ levels \cite{Fulde70,Fulde83,Goremychkin04} --- the inelastic scattering of the conduction electron by the $\Gamma_{1}$$\rightarrow$$\Gamma_{5}$ transition has a strong \textit{quadrupole} matrix element that enhances pair formation and consequently $T_{c}$. This interaction, known as aspherical Coulomb scattering, overcomes the magnetic pair breaking effect due to the \textit{s-f} exchange scattering of the conduction electron by the $\Gamma_{1}$$\rightarrow$$\Gamma_{4}$ transition, which has a strong \textit{dipole} matrix element that suppresses $T_{c}$. The $f$ electrons thus play an important role in skutterudite/heavy-fermion superconductivity via the CEF-split $f$-electron energy levels.

Since it has been shown that PrOs$_{4}$Sb$_{12}$ is a two-band superconductor with different pairing symmetries on each band, we want to find out if LaOs$_{4}$Sb$_{12}$ is a multiband superconductor as well, and if so, whether the symmetries of these multiple gaps are different. Such an investigation of LaOs$_{4}$Sb$_{12}$ will allow us to determine whether the $f$-electrons are necessary for the formation of multiband superconductivity. In general, multiband superconductivity arises when there are more than one conduction band \textit{and} the pair couplings within each band are of different magnitudes.~\cite{Suhl59}. Though filled skutterudites generically contain two hole-like conduction bands, observations of multiband superconductivity were only reported for PrOs$_4$Sb$_{12}$ and PrRu$_4$Sb$_{12}$ --- both containing $f$-electrons.~\cite{Hill08} These $f$-electrons might exert different degrees of influence on the pair coupling within each band, hence giving rise to multiband superconductivity. In this context, it is tempting to attribute multiband superconductivity in Pr-based skutterudites to the presence of $f$-electrons.

There have been evidences that LaOs$_{4}$Sb$_{12}$ is a weak-coupling $s$-wave superconductor. Specific heat measurement showed a discontinuity jump $\Delta C$ at $T_c$ with $\Delta C/ \gamma T_c=1.46$,~\cite{Rotundu06} close to the BCS value of 1.43. Sb-NQR measurement showed a coherence peak at $T_c$ and exponential temperature dependence at low temperatures.~\cite{Kotegawa03} The data can be fitted by an isotropic gap \cite{Kotegawa03,Parker08} with $\Delta(0) \approx 1.6k_B T_c$ or an anisotropic gap $\Delta(\theta)=\delta+(\Delta-\delta)$sin$\theta$ with $\Delta=1.73 k_B T_c$ and $\delta=1.21 k_B T_c$.~\cite{Yogi06} However, to the best of our knowledge, multiband superconductivity has never been reported for this material.

Magnetic penetration depth has been shown to be a valuable tool for probing multiband superconductivity, for instance, in the well-known multiband superconductor MgB$_2$.~\cite{Carrington03} At low temperatures, the deviation $\Delta\lambda(T)$=$\lambda(T)-\lambda(0)$ is sensitive to the low energy excitations of quasiparticles across the superconducting gap. Consequently, the symmetry of the energy gap can be determined from the temperature-dependence of  $\Delta\lambda(T)$. In addition, fits to the superfluid density over the entire temperature range of measurement allows one to detect the existence of multiple gaps as well as their temperature dependences. In this Paper, we report the measurement of magnetic penetration depth $\lambda(T)$ in single crystals of LaOs$_{4}$Sb$_{12}$ down to 85~mK. We found that at low temperatures, $\Delta \lambda(T)$ exhibits an exponential temperature dependence consistent with a $s$-wave gap symmetry. Fitting the data to BCS $s$-wave expression gives a minimum gap value that is significantly smaller than the BCS value, suggestive of multiband superconductivity. The superfluid density data over the entire temperature range are best fitted by a two-band $s$-wave model.

Details of sample growth and characterization are described in Ref.~\onlinecite{Sugawara02}. Measurements were performed utilizing a 21-MHz self-resonant tunnel diode oscillator \cite{Elbert03} with a noise level of 2 parts in 10$^9$ and low drift. The sample was positioned in the center of a induction coil which forms part of the LC resonant tank circuit. The $\Delta\lambda(T)$ is directly proportional to the change in resonant frequency $\Delta f(T)$ of the oscillator, with the proportionality factor $G$ dependent on sample and coil geometries. For a square sample of side $2w$, thickness $2d$, demagnetization $N$, and volume $V$, $G$ is known to vary as $G \propto R_{3D}(1-N)/V$, where $R_{3D}=w/[2(1+(1+2d/w)^2)\arctan(w/2d)-2d/w]$ is the effective sample dimension.~\cite{Prozorov00} For our sample $2w\approx 0.59$~mm and $2d\approx 0.13$~mm. The value of $G$ was determined for a high-purity Al single crystal by fitting the Al data to the extreme non-local expression, and then adjusting for relative sample dimensions. The magnitude of the ac field inside the induction coil was estimated to be less than 40~mOe, and the cryostat was surrounded by a bilayer Mu-metal shield that reduced the dc field to less than 1~mOe. The single crystal sapphire rod which held the sample with GE varnish was attached to the mixing chamber of an Oxford Kelvinox-25 dilution refrigerator to provide cooling. During the experiment, the probing ac field of the inductor coil was directed along the $c$-axis of the sample and the in-plane penetration depth $\lambda_{ab}$ was measured. Since LaOs$_4$Sb$_{12}$ has a cubic crystal structure, we will omit the distinction between axes for the rest of this paper.

As $\lambda(0)\approx$ 4700 \AA~in LaOs$_{4}$Sb$_{12}$,~\cite{Aoki05} it probes a significant depth into the sample and is therefore less sensitive to surface quality, giving result representative of the bulk. At low temperatures ($T \lesssim T_c/3$), it is well established that $\lambda(T)$ of an isotropic $s$-wave superconductor will asymptomatically approach an exponential behavior given by
\begin{equation}
\Delta\lambda(T)=\lambda(0)\sqrt{\frac{\pi\Delta(0)}{2k_{B}T}}\textsf{exp}\left( -\frac{\Delta(0)}{k_{B}T}\right)
\label{eqn:swave}
\end{equation} where $\lambda(0)$ and $\Delta(0)$ are the zero-temperature values of $\lambda$ and the superconducting gap, respectively. For a weakly anisotropic gap or multiple gaps, Eq.~(\ref{eqn:swave}) will still follow but now with $\Delta(0)$ being replaced by the minimum gap value in the system, and $\lambda(0)$ by an effective value which depends on the details of gap anisotropy.

\begin{figure}
\centering
\vspace{-10pt}
\includegraphics[scale=0.3]{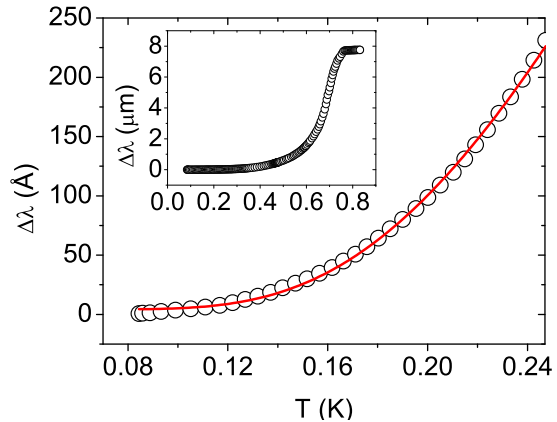}
\vspace{-15pt}
\caption{Temperature dependence of the penetration depth $\Delta\lambda(T)$ in sample 1 at low temperatures. The solid line is a fit to Eq.(~\ref{eqn:swave}). Inset: $\Delta\lambda(T)$ over the entire temperature range.}
\vspace{-15pt}
\label{fig:lambda}
\end{figure}

We measured the $\Delta\lambda(T)$ in two single crystals of LaOs$_4$Sb$_{12}$. An exponential temperature dependence characteristic of a $s$-wave energy gap is observed in the two samples over the temperature range 85~mK--0.26~K. Figure~\ref{fig:lambda} shows the low-temperature variation of $\Delta\lambda(T)$ in sample 1. The inset in Fig.~\ref{fig:lambda} shows $\Delta \lambda (T)$ of sample 1 over the entire temperature range. The $\Delta \lambda (T)$ starts to fall at 0.76~K and reach the transition mid-point at 0.68~K, consistent with the reported value of $T_c$=0.74~K in other experiments.~\cite{Aoki05,Kotegawa03,Andraka10}

The low-temperature data of sample 1 was fitted to Eq.~(\ref{eqn:swave}) up to 0.25~K ($\sim$T$_c$/3), as shown by the solid line in Fig.~\ref{fig:lambda}. The zero-temperature gap value $\Delta (0)$ obtained from the fitting was found to be (0.99$\pm 0.05$)~K or (1.34$\pm 0.07$) $k_B T_c$ (assuming $T_c$=0.74 K), where the error corresponds to fitting the data up to $T=0.23$~K and $T=0.26$~K respectively. For sample 2, fitting the low-temperature data to Eq.~(\ref{eqn:swave}) gives $\Delta(0)=(1.06\pm0.5)$~K or $\Delta(0)=(1.43\pm0.07) k_B T_c$, slightly larger than the value obtained in sample 1. We also fitted the low-temperature data to a power law  $\Delta\lambda(T)\sim T^n$ and obtained $n$=4, excluding the possibility of line nodes in the energy gap which gives $\Delta\lambda(T)\sim T$ or $\Delta\lambda(T)\sim T^2$ in the presence of impurities. As sample 1 has a sharper transition near $T_c$, we will focus on it for the rest of this paper.  The value $\Delta(0)=1.34 k_B T_c$ is significantly smaller than the weak-coupling BCS value of 1.76$k_B T_c$. This suggests the possibility of anisotropic gap or multiple gaps (multiband superconductivity).

\begin{figure}
\centering
\vspace{-10pt}
\includegraphics[scale=0.3]{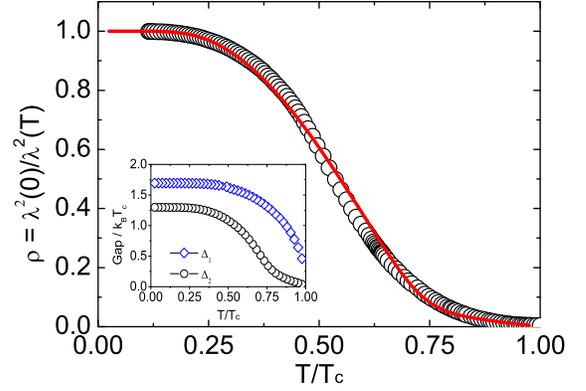}
\vspace{-15pt}
\caption {Experimental data (O) and theoretical fit (solid line) for the normalized superfluid density $\rho(T)$. Insert: temperature dependence of the two energy gaps from the theoretical fit.}
\vspace{-15pt}
\label{fig:rho}
\end{figure}

The scenario of multiple gaps is more likely, since in addition, we also observed an unusually long suppression of the normalized superfluid density $\rho(T)=\lambda^2(0)/\lambda^2(T)$ near $T_c$ which could not be due to gap anisotropy (Fig.~\ref{fig:rho}). On the contrary, the presence of multiple gaps often results in suppression of $\rho(T)$ near $T_c$ due to the presence of the smaller superconducting gap.~\cite{Xiang96,Luo05} This feature should be intrinsic to the material since it has been observed in another two samples and could not have been caused by some magnetic ordering, as $\mu$SR experiment observed no spontaneous magnetism even down to 20~mK.~\cite{Aoki05}

We propose a two-band model to fit our data. In this model, the Hamiltonian is given by
\begin{eqnarray}
H &=&\sum_{ik\sigma }\epsilon _{i,k}c_{i,k\sigma }^{+}c_{i,k\sigma
}+\sum_{ikk^{\prime }}V_{i,kk^{\prime }}c_{i,k^{\prime }\uparrow
}^{+}c_{i,-k^{\prime }\downarrow }^{+}c_{i,-k\downarrow }c_{i,k\uparrow
}\;\;\;\;\nonumber \\
&&+\sum_{kk^{\prime }}V_{3,kk^{\prime }}c_{1,k^{\prime }\uparrow
}^{+}c_{1,-k^{\prime }\downarrow }^{+}c_{2,-k\downarrow }c_{2,k\uparrow
}+H.c.\hspace{2cm}
\end{eqnarray}where $i$=1,2 represents the first and second bands. $c_{1,k\sigma}$ and $c_{2,k\sigma}$ are the corresponding electron operator.  $V_{1,kk^{\prime}}=V_1\gamma_{1,k}\gamma_{1,k^{\prime}}$ and $V_{2,kk^{\prime}}=V_2\gamma_{2,k}\gamma_{2,k^{\prime}}$ are the reduced pair coupling for the two bands. $V_{3,kk^{\prime}}=V_3\gamma_{1,k}\gamma_{2,k^{\prime}}$ is the interband pair coupling. This model has been used to describe the two-band superconductor MgB$_2$.~\cite{Nakai02}

Taking the BCS mean-field approximation, the interaction between the two band is decoupled as
\begin{eqnarray}
H &\approx &\sum_{ik\sigma }\epsilon _{i,k}c_{i,k\sigma }^{+}c_{i,k\sigma} +\nonumber\\
&&\left( \Delta _{i}\sum_{k}\gamma _{i,k}c_{i,k\uparrow
}^{+}c_{i,-k\downarrow }^{+}+\Delta _{i}^{\ast }\sum_{k}\gamma
_{k}c_{i,-k\downarrow }c_{i,k\uparrow }\right)
\end{eqnarray} where $\Delta _{1} =\sum_{k}\left( V_{1}\gamma _{1,k}\langle c_{1,-k\downarrow}c_{1,k\uparrow }\rangle +V_{3}\gamma _{2,k}\langle c_{2,-k\downarrow}c_{2,k\uparrow }\rangle \right)$ and $\Delta _{2} =\sum_{k}\left( V_{2}\gamma _{2,k}\langle c_{2,-k\downarrow}c_{2,k\uparrow }\rangle +V_{3}\gamma _{1,k}\langle c_{1,-k\downarrow}c_{1,k\uparrow}\rangle \right)$.

We propose the tight-binding band dispersion \begin{eqnarray}
\epsilon _{i,k} &=&-2t_{i}\left( \cos k_{x}+\cos k_{y}+\cos k_{z}\right) \nonumber\\
&&+4t_{i}^{\prime }\left(\cos k_{x}\cos k_{y}+\cos k_{x}\cos k_{z}+\cos k_{y}\cos k_{z}\right)\nonumber\\
&&-\mu_{i}.
\end{eqnarray} Here $t_{i} (i=1,2)$ are the hopping constants, $t_{i}^{\prime }$ are the next nearest-neighbor hoppings, and $\mu _{i}$ are the chemical potentials of the two bands. The same tight-binding band dispersion with a slight modification due to antiferromagnetism has been used to describe multiband superconductivity in the electron-doped cuprate superconductor Pr$_{2-x}$Ce$_x$CuO$_4$.~\cite{Luo05}

We next solve, self-consistently, a set of equations of $\Delta _{1},\Delta _{2},\mu _{1},\mu _{2}$ obtained by $\delta F/\delta x=0$
$(x=\Delta _{1},\Delta _{2},\mu _{1},\mu _{2})$, where $F =-\frac{1}{\beta }\ln Z=-\frac{1}{\beta }\ln e^{-\beta H}$ is the free energy, for given model parameters
 $t_{i},t_{i}^{\prime },V_{i},V_{3},n_{1},n_{2}$. The
superfluid density along a certain direction, such as $x$-direction, is given by
\begin{eqnarray*}
\rho _{s}^{x} &=&\sum_{i=1,2}\sum_{k}\left( \langle \frac{\partial
^{2}\epsilon _{i,k}}{\partial k_{x}^{2}}\rangle +2\left( \frac{\partial
\epsilon _{i,k}}{\partial k_{x}}\right) \frac{\partial f\left(
E_{i,k}\right) }{\partial E_{i,k}}\right)
\end{eqnarray*} where $f\left(E_{i,k}\right)$ is the Fermi function.
In our fitting, we assume two bands to have $s$-wave gaps, namely, $\gamma _{1,k}$=$\gamma_{2,k}$=1. The choice of gap symmetry is supported by the low-temperature exponential behavior in $\Delta \lambda$ (this work) and Sb-NQR experiment.~\cite{Kotegawa03} For LaOs$_4$Sb$_{12}$, the model parameters used are $\left( n_{1,}t_{1},t_{1}^{\prime
},V_{1},n_{2,}t_{2},t_{2}^{\prime },V_{2},V_{3}\right) =\left(0.08,t_2/4.4,-0.25t_{2},0.08t_2,0.25,t_2,-0.3t_{2}, 0.103t_{2}, 0.00\right.$
$\left. 08t_{2}\right).$ Here the hopping integral $t_{2}=1$ is taken as the unit of energy in the whole
fitting, and is related to critical temperature as $T_c \approx 0.049t_{2}$.

From the fitting, we obtained $\Delta_1 (0)=1.69k_BT_c$, $\Delta_2 (0)=1.30k_BT_c$, and $n_2/n_1=3.13$. The calculated value of $\Delta_2 (0)=1.30k_BT_c$ agrees well with the value of the zero-temperature gap value $\Delta(0)=(1.34\pm 0.07)k_B T_c$ obtained from low-temperature data of $\Delta\lambda(T)$. The two values agree, since at low temperatures, the smaller energy gap will be the effective gap for quasiparticle excitation. The calculated value of $\Delta_1 (0)=1.69 k_B T_c$ is also consistent with the weak-coupling value of $1.76 k_B T_c$.~\cite{Kotegawa03,Rotundu06} Figure~\ref{fig:rho} shows that the experimental data of $\rho(T)$, in particular the long suppression near $T_c$, can be well fitted by our two-band model. The long suppression is effectively caused by the smaller gap $\Delta_2$ vanishing at higher temperatures due to interband coupling with the bigger gap $\Delta_1$ (Inset of Fig.~\ref{fig:rho}).

Our present finding reveals a two-band $s$-wave superconductivity in LaOs$_4$Sb$_{12}$. Given the close Fermi topologies of PrOs$_4$Sb$_{12}$ and LaOs$_4$Sb$_{12}$, the present result shows that multiband superconductivity of PrOs$_4$Sb$_{12}$ persist even when the $f$-electrons are removed. This excludes the $f$-electrons from being the origin of multiband superconductivity in PrOs$_4$Sb$_{12}$. Though the $f$-electrons are not responsible for the multiband superconductivity, they may however influence the behavior of each band. Our present finding of $s$-wave symmetry in both bands suggests that the two bands evolve differently in the presence of $f$-electrons, namely into a nodal band and a fully-gapped band in PrOs$_4$Sb$_{12}$. This unconventional effect of $f$-electrons in PrOs$_4$Sb$_{12}$ may be related to the weakly split CEF levels,~\cite{Bauer02,Maple02,Goremychkin04} since it was found that the nodal feature disappears once PrOs$_4$Sb$_{12}$ is doped with Ru.~\cite{Elbert04,Elbert05} We therefore suggest that theoretical models for PrOs$_4$Sb$_{12}$ should take the multiband superconductivity of LaOs$_4$Sb$_{12}$ as a starting point, and develop the subsequent modifications in superconducting properties due to the adding of $f$-electrons.

The magnetic penetration depth data of PrOs$_4$Sb$_{12}$ and PrRu$_4$Sb$_{12}$ had been measured by one of us with the same experimental technique employed here.~\cite{Elbert03,Elbert04,Elbert05} We tried to fit the data of PrOs$_4$Sb$_{12}$ with a similar two-band $s$-wave model, but was not successful --- this is consistent with the unconventional (point node) nature of one of its gaps. Also, we were able to fit the PrRu$_4$Sb$_{12}$ data using just one superconducting gap, showing that the multiband superconductivity of PrRu$_4$Sb$_{12}$, as shown in thermal conductivity data, does not significantly affect the magnetic penetration depth data.

In summary, we report measurements of the magnetic penetration depth $\lambda(T)$ in single crystals of LaOs$_4$Sb$_{12}$ down to 85~mK. We find that $\lambda(T)$ exhibits an exponential temperature dependence at low temperatures with the zero-temperature gap value $ \Delta (0)= (1.34 \pm 0.07) k_B T_c$. Our results show that LaOs$_4$Sb$_{12}$ is a two-band $s$-wave superconductor. Given the close Fermi topology of PrOs$_4$Sb$_{12}$ and LaOs$_4$Sb$_{12}$, this suggests that the $f$-electrons are not the origin of multiband superconductivity in these two materials.

This work was supported by the NSF through Grant No. DMR99-72087 and DMR01-07253. Some of the work was carried out in the Materials Research Laboratory and Center for Microanalysis of Materials, University of Illinois, which is partially supported by the U.S. Department of Energy under grant DEFG02-91-ER45439. We acknowledge financial support by National Research Foundation of Singapore. This work was also supported partially by NSFC (Nos.10934008, 10874215, and 11174115) and MOST 973 Project (No. 2011CB309703) of China, and Grant-in-Aid for Scientific Research on Innovative Areas ``Heavy Electrons" (No. 21102503) of The Ministry of Education, Culture, Sports, Science, and Technology, Japan.

\bibliography{LaOsSb}

\end{document}